\documentclass[12pt,preprint]{aastex}
\usepackage{epsfig}

\newcommand{\degrees}{$^\circ$}
\newcommand{\kms}{$\mbox{km~s}^{-1}$}
\newcommand{\myr}{$\mbox{mas~yr}^{-1}$}
\newcommand{\arcs}{$^{\prime \prime}$}

\begin{document}

\shorttitle{Vela Pulsar's proper motion and parallax}
\shortauthors{Dodson et al.}

\title{The Vela Pulsar's proper motion and parallax derived from VLBI
observations}

\author{R. Dodson\altaffilmark{1,3} 
	D. Legge\altaffilmark{1} 
	J.E. Reynolds\altaffilmark{2} and
	P.M. McCulloch\altaffilmark{1}}

\affil{currently at ISAS, Japan}
\altaffiltext{1}{School of Mathematics and Physics, University of Tasmania, 
     Australia}
\altaffiltext{2}{Parkes Observatory, PO Box 276, Parkes, NSW 2870, Australia}
\altaffiltext{3}{ISAS, Yoshinodai 3-1-1 , Sagamihara, Kanagawa
229-8510, Japan}

\email{rdodson@vsop.isas.ac.jp}


\begin{abstract}

The Vela pulsar is the brightest pulsar at radio wavelengths. It was
the object that told us (via its glitching) that pulsars were solid
rotating bodies not oscillating ones. Along with the Crab pulsar is it
the source of many of the models of pulsar behavior. Therefore it is
of vital importance to know how far away it is, and its origin.

The proper motion and parallax for the Vela pulsar have been derived
from 2.3 and 8.4 GHz Very Long Baseline Interferometry (VLBI)
observations. The data spans 6.8 years and consists of eleven epochs.
We find a proper motion of $\mu_{\alpha {\rm cos}\delta}= -49.61 \pm
0.06,\ \mu_\delta= 29.8 \pm 0.1$ \myr\ and a parallax of $3.4 \pm 0.2$
mas, which is equivalent to a distance of $293_{-17}^{+19}$ pc. When
we subtract out the galactic rotation and solar peculiar velocity we
find $\mu_* = 45 \pm 1.3$ \myr\ with a position angle (PA) of
$301^\circ\pm1.8$ which implies that the proper motion has a small but
significant offset from the X-ray nebula's symmetry axis.


\end{abstract}


\section{Introduction}

The Vela pulsar is an archetypal young pulsar. It was the first to
be observed to glitch \citep{rad_glitch} and was always associated with
the large Vela supernova remnant (SNR). Since Vela is a young pulsar
and displays a significant amount of glitching behavior and timing
noise \citep{vela_timing}, no accurate proper motion measurements from
timing observations have been possible; it is only through direct
proper motion observations that we can trace its path on the sky. This
is true of nearly all young pulsars, which are usually the most
interesting. These have, for example, the possibility of discovering
the associated birth SNR.
%
The first accurate measurements of Vela's proper motion were produced
using the Parkes-Tidbinbilla Interferometer (PTI) \citep{bailes}. This
275 km baseline at 1.6~GHz gave a resolution of 140 mas. This
provided the first accurate measurement of the proper motion and thus,
by extrapolation, the birth site. We have extended this work to allow
the third dimension to be derived via the parallax. Measurement of
relative motion on milli-arcsecond scales is a challenging task and
several attempts have been made, with various success. Previous
observations have been made with ground based optical observations,
with connected radio interferometers (PTI and the VLA) and the Hubble
Space Telescope (HST). We now present results from the Australian Long
baseline array (LBA).
 
We have 18 observations of the position and can provide the most
accurate measurements of parallax and proper motion. Recent
measurements of the position of Vela against a number of background
sources has been reported by \citet{deluca_pm,caraveo_pm}. Our more
accurate values, using completely independent methods, are compared
with theirs, confirming the parallax values. Differences in the
derivation of the proper motion are discussed.  We find good agreement
if we account for the probable galactic rotation of the reference
stars.

The distance to the Vela pulsar was originally estimated from the SNR
distance by comparison to the Cygnus loop and IC443. This lead to an
estimate of 500 pc \citep{milne}, however it has long been argued
that this is an overestimate.
Our results have been foreshadowed by the predictions in several
papers that have made the case for the distance to the Vela pulsar to
be drastically reduced. Analysis of X-ray observations of the pulsar
made with ROSAT \citep{page} and Chandra \citep{pavlov}, of the SNR
also with ROSAT \citep{bocchino} and with optical absorption
\citep{jenkins_wallerstein,cha} have all suggested that the original
distance estimate should perhaps be halved. All of these results,
however, are to some extent model dependent and therefore doubt has
existed over their accuracy.


Furthermore, our results allow the testing of the prediction of
\citet{spruit_phinney} on the alignment of the spin axes of pulsars
with their proper motion vector. The spin axis can be derived from the
high resolution images from Chandra. These allow the symmetry, and
thus presumably the spin, axes to be directly discerned. Only two
cases have been tested so far, the Crab \citep{crab_hst} and Vela. The
extra accuracy we can provide refines the conclusion reached by
\citet{pavlov_vela_image,helfand} for the Vela alignment.
A note of caution should be raised in that an alternate explanation
for the X-ray nebula around Vela has been put forward \citep{desh_rad}
which explains the structure in terms of the particle beam from the
polar cap. This description requires that only the projected spin axis
lies along the proper motion axis.
If the model of \citet{spruit_phinney} is correct any misalignment of
the axes allows the us to estimate the timescale of the impulse in
terms of the initial rotation period, a value of great importance to
understanding the core collapse. The accuracy, therefore, of the
alignment of the axes is of great theoretical importance.

\section{Observations and Data reduction}

The LBA array is an Australian national facility and is usually made
up of six telescopes, three operated by the CSIRO (Australian
Telescope Compact Array, Parkes and Mopra), two by the University of
Tasmania (Hobart and Ceduna) and one in South Africa operated by the
Hartebeesthoek Radio Astronomical observatory. In addition DSN
telescopes at Tidbinbilla are often included. The observations
reported here are between the Tidbinbilla 70m and the Hobart 26m
antennae, a baseline of 832 km and at two 16-MHz wide frequencies;
2.29 and 8.425-GHz. In addition there was one observation with the
Australian array, 32-MHz centered on 8.417-GHz, although only the Hobart to
Tidbinbilla data was directly used for the results in this
paper. Details of all the telescopes and their parameters are
available via the Australian Telescope National Facility (ATNF)
website\footnote{http://www.atnf.csiro.au/vlbi}.  All data were
recorded using beam-switching on a $\sim$5 minute cycle, using the
extra galactic phase-reference source Vela-G ($\alpha_{J2000} =
08^h33^m22^s.31563$, $\delta_{J2000} = -44^\circ41^\prime38^{\prime
\prime}.71463$). Vela-G is part of the Radio Reference Frame and was
an International Celestial Reference Frame (ICRF) candidate source
\citep{ma_98}.

The first three epochs observations of the Vela pulsar (Table
\ref{tab:observations}) were made using MkIII/MkIIIA VLBI recording
systems, using recording mode B, which provides 14 contiguous
frequency channels each 2 MHz wide in right-hand circular polarization
at 2.3~GHz. Sampling was one-bit (two-level) for all data. None of the
Mark III/IIIA observations used the pulsar gating nor binning described
later. Only the MkIII positions were available for this analysis.
The MkIII/MkIIIA VLBI data were processed at the Washington Mark IIIA
correlator, located at the U.S. Naval Observatory (USNO). Phase models
were applied with the CALC 6.0 package, supplying the best possible
values of station locations, clock models and earth orientation
parameters.  The raw data were then exported in "FRNGX" format and
further analyzed following the phase-referencing technique described
by Lestrade et al.  (1990), and used the software SPRINT, developed by
J.-F. Lestrade.


The S2 recordings system and correlator for LBA were commissioned in
the mid-1990s and observations were transferred to this system.
In all observations, two separate frequency channels of 16 MHz
bandwidth were recorded with two-bit (four-level) sampling. Most of
the observations of the Vela pulsar using the S2 system were made with
one 16 MHz channel at 2.3 GHz (13 cm) and one at 8.4 GHz (3 cm),
taking advantage of the dual 13/3 cm (right-circular polarization)
receivers at both Hobart and Tidbinbilla. In the final epoch, using
the whole LBA at 8.4~GHz, two 16 MHz channels were used to form one
contiguous 32 MHz band.

The S2 correlator is operated by ATNF at its Marsfield headquarters in
Sydney. This wonderfully flexible device can pulsar bin (as opposed to
pulsar gate) with thirty-two bins across the Vela pulsar's period,
thereby limiting the dispersion to individual channels rather than the
entire band. There is, however, no significant dispersion at these
observing frequencies for Vela.  We used only those pulsar bins
containing significant flux, usually one or two out of thirty two,
gaining up to a factor of 5.6 in signal to noise.


Post correlation the data were fringe fitted at the nominal pulsar
position using the binned flux, with ten minute solution interval in
{\small AIPS}\footnote{AIPS, Astronomical Image Processing System,
developed and maintained by the NRAO.}, and then exported. Further
processing was done with {\small DIFMAP} \citep{difmap}. The data were
flagged and averaged to generate a statistical weight. The offsets of
Vela-G from its nominal reference position were found by fitting a
point source to the visibilities. The position of Vela was shifted
(phase rotated) to correct for these offsets. We then selfcalibrated
the Vela phases in {\small AIPS} at 5 minutes intervals, and copied
those corrections to the Vela-G visibilities. The final offsets of
Vela-G from the reference position, all very small, were used to
calculate the final Vela position.

Lobe ambiguity was not an issue even though this was a single baseline
experiment, as observing with two widely separated frequencies broke
any degeneracy. The final observation, which was with two 8.4-GHz
bandpasses, had multiple baselines with which to identify the correct
lobe. Only the Tidbinbilla to Hobart baseline was used for
measurements of the positions.

\section{Results}

We find a pulsar position of $\alpha_{J2000} = 08^h35^m20^s.61153 \pm
0.00002$, $\delta_{J2000} = -45^\circ10^\prime34^{\prime \prime}.8755
\pm 0.0003$ for a reference epoch of 2000.0. The proper motion is
$\mu_{\alpha {\rm cos}\delta} = -49.61 \pm 0.06,\ \mu_\delta = 29.8
\pm 0.1$ \myr\ and a parallax of $3.4 \pm 0.2$ mas, equivalent to a
distance of $293_{-17}^{+19}$ pc.


\subsection{Sources of error in phase referenced observations}
\label{Sources of Error}

\subsubsection{Source structure of the phase reference source}

Whilst strictly the pulsar itself is the phase-reference in this
experiment (with a binned flux of 5 and 0.6 Jy at 2.3 and 8.4 GHz) the
effect of significant structure is the same in either case. If the
source is non-symmetrical the incoming wavefront arriving at the
baseline will not have the assumed intrinsic zero phase, degrading the
source fitting. Therefore for our final observation we used the full
array to allow imaging of both Vela-G and the pulsar. As we were looking
for local structure we were able to use selfcalibration.

Following the Chandra observations which unveiled the fine X-ray
structure around the pulsar we carefully searched for radio structure
at VLBI resolutions in the off pulse data (calibrated with the on
pulse phases). We found no discernible structure on baselines between
2.6 and 15.6 M$\lambda$ (equivalent to a full width half maximum of 16
-- 95 mas) within 10\arcs\ of the pulsar. The peak point source flux
was 0.8~mJy/beam in line with expectations given the RMS of
0.1~mJy. This is of no great surprise as the radio structure around
the pulsar has been reported to resolve out at resolutions finer than
a few arcseconds \citep{dodson_vrn,lewis_02}.

As pulsar radiospheres have angular sizes much less than 1 mas, they
may be treated as a point source. (Scattering caused by the ISM may
cause observable angular broadening of the pulsar image but is
unlikely to bias the observed position significantly).
The phase-reference source has not been optically identified but is
most likely a distant AGN, and might well have source structure that
produces systematic errors in the positions determined at each epoch for
the pulsar. The fact that in most observations we used the Vela pulsar as
the reference source for the weaker Vela G does makes no essential
difference in this respect. With this in mind, the final epoch of our
observations was used to image the reference source with the full LBA
array.
No structure brighter than 1~mJy was found around Vela-G (18
mJy) at 8.4 GHz, with an image RMS of 0.3~mJy. The phase residuals had
a RMS of 4\degrees .
We note here that Vela-G is a well known extra-galactic source and
considered for the ICRF (candidate source ICRF J083322.3-444138)
\citep{ma_98}.


\subsubsection{Source structure of the ionospheric delay}

A considerable phase error can be caused by the different ionospheric
delay encountered over the different antennae. Several approaches
exist to address these problems, the best being using multiple
frequencies in the VLBI observations and solving for the ionosphere as
part of the reduction \citep{brisken_apj_0950}. Unfortunately the S2
system used in our experiments does not have enough spanned bandwidth
to allow this approach to be used. We therefore we have looked into
using the measured Total Electron Count (TEC) from GPS observations
\citep{vla_memo_23}. However the best data is on a 5\degrees\
grid. The separation of our phase-reference and the target source is
0.7\degrees\ and therefore GPS data cannot provide a useful
correction.

As no correction was possible we modeled the data quality using
DIFWRAP \citep{lovell_difwrap} which allows an error estimate that
includes all possible contributions without attempting to identify
them. This approach involves exploring a range of model parameters to
identifying the range of acceptable fits, which should be the bounds
of the multi-dimensional one sigma contour. The errors found are
indeed approximately equivalent to one sigma, as confirmed by fact
that the proper motion fit has a reduced $\chi^2$ of 1.1.

\subsection{Space velocity}
\label{Space velocity}

Two corrections need to be applied to our results to the true local
motion of the Vela pulsar in its local environment. Our observations
are directly tied to the ICRF, therefore the solar peculiar motion and
the galactic rotation contribute to the observed proper motion, and
need to be removed. We have used the solar constants from
\citet{dehnen_binney} of $10\pm0.36,5.25\pm0.62,7.17\pm0.38$ \kms\ in
galactic coordinates. 
We have used a flat rotation curve ($\Omega_0=220$ \kms , $R_0=8.5$
kpc; \citet{fich_rot}), which produces a local proper motion of
$-5.4~\mu_l$ \myr . This compares with $-5.7~\mu_l$ \myr\ from the
local values for the Oort constants, as found by
\citet{feast_whitelock}. The corrections are shown in the Table
\ref{tab:corr}.
The dominant source of error is from the uncertainties in the solar
peculiar motion parameters. We have used the measured uncertainties in
our observations and combined those with the models. We have ignored
the possibilities of systematic errors or alternate models. This gives
us an angular motion, at Vela's local standard of rest, of $\mu_\alpha
= -38.5 \pm 1.2$ \myr , $\mu_\delta = 23 \pm 1.5$ \myr\ or $\mu_* =
45\pm 1.3$ \myr\ at a position angle of $301^\circ \pm 1.8$.  The
pulsars' transverse space velocity is therefore $62 \pm 2$ \kms . The
PA no longer lies quite along the spin axis (e.g. $310^\circ \pm 1.5$
\citet{helfand}, or $307^\circ \pm 2$ \citet{pavlov_vari}), which may
strengthen the case that the impulse timescale was not quite long
enough to average the off-axis component to zero.




\section{Discussion}
\subsection{Fitting methods}

We used the publicly available proper motion fitting routines, {\small
PMPAR}\footnote{http://nacho.princeton.edu/\~walterfb/pmpar/pmpar.html},
created by W. Brisken. 
As each epoch had quite different observation spans and phase
stability great care was required to ensure that the error estimates
were accurate. The traditional approach has been to assume that the
errors are a fraction of the beam size, but these ignore variations in
the observing conditions during the experiment. 
We used {\small DIFWRAP} \citep{lovell_difwrap} to measure the
complete range of errors and derived error estimates which are
realistic. Where we only had the archival positions, and not the data,
we have used the median value of the difference between the formal
errors and the errors found with {\small DIFWRAP}. It is particularly
important to get the errors correct, as the effect to be measured is
small and the variation between the data quality is large.

\citet{deluca_pm} point out that, as they calculate the proper motion
using data collected on nearly identical day-numbers, their proper
motion measurement is not contaminated by the parallax. They refer to
this as a `pure' proper motion. This is a concern with very short time
baselines, which would blend the parallax and proper motion. We,
however, have data spanning seven years and the correlation between
the parameters is low.

\subsection{Comparison of proper motion with other studies}
\label{Comparison of proper motion with other studies}

\begin{figure}
\begin{center}
\epsfig{file=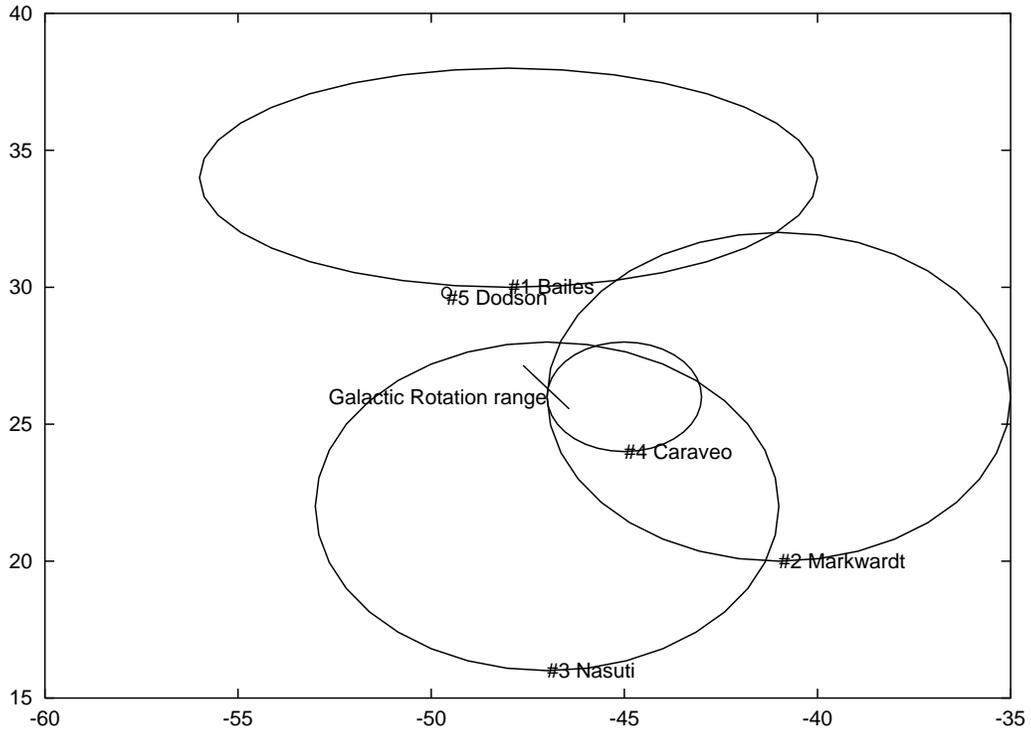,width=10cm,angle=270}
\caption{Vela proper motion determinations in RA and Declination (mas
per year) with 2-$\sigma$ error ellipses. These are labelled with the
reference name and row number from Table \ref{tab:corr}. The effect of
adding the correction for the galactic rotation of calibrators for
distances between 2 and 10kpc is shown for the our result.}
\label{fig:errors}
\end{center}
\end{figure}

Our calculated proper motion and parallax can also be compared with
recent values obtained by optical proper motion studies and
phase-referencing VLBI. We ignore the historic observations, which
were blighted by poor resolution and low elevation, and concentrate on
the two Radio VLBI observations, this one and \citet{bailes} and the
four optical observations, one purely ground based \citep{nasuti}, one
ground and HST \citep{markwardt} and two reports from the HST data set
\citep{deluca_pm,caraveo_pm} for which we take only the latest
results. The results found by each of these studies is shown in
Table~\ref{table:Vela proper motion}. The table shows the proper
motions in right ascension and declination for the pulsar along with
the reference for this work. The proper motions listed do not account
for the rotation of the Galaxy nor the peculiar motion of the Sun in
the local galactic potential being only with reference to the calibrators
used. The radio reference is an extra-galactic source, and the optical
references are a significant number of field stars. Both styles of
observations are internally consistent to within two standard
deviations (see Figure \ref{fig:errors}), with improvement of errors
over time. With the reduction of errors, however, the radio and
optical results are steadily becoming less compatible.

The major advantage our observation have over the HST ones (other than
the formal resolution being approximately one hundred times better) is
that the reference is tied to a well defined reference frame. For the
optical observations only field stars were available. The distance to
the reference stars in the HST must be significantly greater than that
of the pulsar, otherwise they would have had observable parallaxes
themselves. Not knowing a distance we have assumed that they lie
between two and ten kpc and calculated what the apparent proper motion
would be for these limits using the standard flat rotation curve. We
find that the galactic rotation contributes between -5 -- -3 \myr\ to
the proper motion in $l$. As all the sources would have proper motions
within this range the scatter is within the HST errors of 1 mas. This
contribution was not included in the calculations of
\citet{caraveo_pm}, and when it is the two sets of observations are
consistent. Figure \ref{fig:errors} includes a line representing the
range that having the reference source at two to ten kpc would have
contributed to our result. This means that the radio results are now
consistent with all the optical results. This contribution, of course,
needs to be removed to produce the correct space velocity and position
angle for the pulsar.

\section{Conclusions}

We have measured the proper motion and parallax of the Vela pulsar to
an unprecedented accuracy ($\mu_{\alpha {\rm cos}\delta}= -49.61 \pm
0.06,\ \mu_\delta= 29.8 \pm 0.1$ \myr\ , $\pi = 3.4 \pm 0.2$ mas), and
have been able to convert these back to the space velocity and
position angle of the pulsar in its local environment with greater
precision that previously possible ($62 \pm 2$ \kms at
$301^\circ\pm1.8$), because of the unambiguity in the radio reference
frame. This allows the precise comparison of the Vela X-ray nebula
symmetry axis and the proper motion of the Vela pulsar, opening
insights into the timescale of the core collapse processes.

\section{Acknowledgments} 
The Long Baseline Array is part of the Australia Telescope, funded by
the Commonwealth of Australia and operated by ATNF and the University
of Tasmania as a National Facility. This research has made use of the
SIMBAD database operated at CDS, Strasbourg, France. This research has
made use of NASA's Astrophysics Data System Abstract Service. The
author would like to expressly thank Warwick Wilson, head of
Engineering for the ATNF, for always being prepared to listen to even
the most outlandish correlator configuration request and often
implementing them. Dr Brisken provided extremely helpful remarks and
comments, as did the referee, Dr Caraveo, and we thank both.
%

\begin{thebibliography}{29}
\expandafter\ifx\csname natexlab\endcsname\relax\def\natexlab#1{#1}\fi

\bibitem[{{Bailes} {et~al.}(1989){Bailes}, {Manchester}, {Kesteven}, {Norris},
  \& {Reynolds}}]{bailes}
{Bailes}, M., {Manchester}, R.~N., {Kesteven}, M.~J., {Norris}, R.~P., \&
  {Reynolds}, J. 1989, \apjl, 343, L53

\bibitem[{{Bocchino} {et~al.}(1999){Bocchino}, {Maggio}, \&
  {Sciortino}}]{bocchino}
{Bocchino}, F., {Maggio}, A., \& {Sciortino}, S. 1999, \aap, 342, 839

\bibitem[{{Brisken} {et~al.}(2000){Brisken}, {Benson}, {Beasley}, {Fomalont},
  {Goss}, \& {Thorsett}}]{brisken_apj_0950}
{Brisken}, W.~F., {Benson}, J.~M., {Beasley}, A.~J., {Fomalont}, E.~B., {Goss},
  W.~M., \& {Thorsett}, S.~E. 2000, \apj, 541, 959

\bibitem[{{Caraveo} {et~al.}(2001){Caraveo}, {De Luca}, {Mignani}, \&
  {Bignami}}]{caraveo_pm}
{Caraveo}, P.~A., {De Luca}, A., {Mignani}, R.~P., \& {Bignami}, G.~F. 2001,
  \apj, 561, 930

\bibitem[{{Caraveo} \& {Mignani}(1999)}]{crab_hst}
{Caraveo}, P.~A. \& {Mignani}, R.~P. 1999, \aap, 344, 367

\bibitem[{{Cha} {et~al.}(1999){Cha}, {Sembach}, \& {Danks}}]{cha}
{Cha}, A.~N., {Sembach}, K.~R., \& {Danks}, A.~C. 1999, \apjl, 515, L25

\bibitem[{{Cordes} {et~al.}(1988){Cordes}, {Downs}, \&
  {Krause-Polstorff}}]{vela_timing}
{Cordes}, J.~M., {Downs}, G.~S., \& {Krause-Polstorff}, J. 1988, \apj, 330, 847

\bibitem[{{De Luca} {et~al.}(2000){De Luca}, {Mignani}, \&
  {Caraveo}}]{deluca_pm}
{De Luca}, A., {Mignani}, R.~P., \& {Caraveo}, P.~A. 2000, \aap, 354, 1011

\bibitem[{{Dehnen} \& {Binney}(1998)}]{dehnen_binney}
{Dehnen}, W. \& {Binney}, J.~J. 1998, \mnras, 298, 387

\bibitem[{Dodson {et~al.}(2003 Accepted)Dodson, Lewis, McConnell, \&
  Deshpande}]{dodson_vrn}
Dodson, R., Lewis, D., McConnell, D., \& Deshpande, A. 2003 Accepted, \mnras

\bibitem[{{Feast} \& {Whitelock}(1997)}]{feast_whitelock}
{Feast}, M. \& {Whitelock}, P. 1997, \mnras, 291, 683

\bibitem[{Fich {et~al.}(1989)Fich, Blitz, \& Stark}]{fich_rot}
Fich, M., Blitz, L., \& Stark, A. 1989, \apj, 342, 272

\bibitem[{{Helfand} {et~al.}(2001){Helfand}, {Gotthelf}, \&
  {Halpern}}]{helfand}
{Helfand}, D.~J., {Gotthelf}, E.~V., \& {Halpern}, J.~P. 2001, \apj, 556, 380

\bibitem[{{Jenkins} \& {Wallerstein}(1995)}]{jenkins_wallerstein}
{Jenkins}, E.~B. \& {Wallerstein}, G. 1995, \apj, 440, 227

\bibitem[{{Lewis} {et~al.}(2002){Lewis}, {Dodson}, {McConnell}, \&
  {Deshpande}}]{lewis_02}
{Lewis}, D., {Dodson}, R., {McConnell}, D., \& {Deshpande}, A. 2002, in ASP
  Conf. Ser. 271: Neutron Stars in Supernova Remnants, 191

\bibitem[{{Lovell}(2000)}]{lovell_difwrap}
{Lovell}, J. 2000, in Astrophysical Phenomena Revealed by Space VLBI, 2000,
  Eds.: H. Hirabayashi, P.G. Edwards, and D.W. Murphy, Published by the
  Institute of Space and Astronautical Science, 301--304

\bibitem[{{Ma} {et~al.}(1998){Ma}, {Arias}, {Eubanks}, {Fey}, {Gontier},
  {Jacobs}, {Sovers}, {Archinal}, \& {Charlot}}]{ma_98}
{Ma}, C., {Arias}, E.~F., {Eubanks}, T.~M., {Fey}, A.~L., {Gontier}, A.-M.,
  {Jacobs}, C.~S., {Sovers}, O.~J., {Archinal}, B.~A., \& {Charlot}, P. 1998,
  \aj, 116, 516

\bibitem[{{Markwardt} \& {Ogelman}(1994)}]{markwardt}
{Markwardt}, C.~B. \& {Ogelman}, H.~B. 1994, Bulletin of the American
  Astronomical Society, 26, 871

\bibitem[{{Milne}(1968)}]{milne}
{Milne}, D.~K. 1968, Proceedings of the Astronomical Society of Australia, 1,
  93

\bibitem[{{Nasuti} {et~al.}(1997){Nasuti}, {Mignani}, {Caraveo}, \&
  {Bignami}}]{nasuti}
{Nasuti}, F.~P., {Mignani}, R., {Caraveo}, P.~A., \& {Bignami}, G.~F. 1997,
  \aap, 323, 839

\bibitem[{{Page} {et~al.}(1996){Page}, {Shibanov}, \& {Zavlin}}]{page}
{Page}, D., {Shibanov}, Y.~A., \& {Zavlin}, V.~E. 1996, in Roentgenstrahlung
  from the Universe, 173--174

\bibitem[{{Pavlov} {et~al.}(2001{\natexlab{a}}){Pavlov}, {Kargaltsev},
  {Sanwal}, \& {Garmire}}]{pavlov_vari}
{Pavlov}, G.~G., {Kargaltsev}, O.~Y., {Sanwal}, D., \& {Garmire}, G.~P.
  2001{\natexlab{a}}, \apjl, 554, L189

\bibitem[{{Pavlov} {et~al.}(2000){Pavlov}, {Sanwal}, {Garmire}, {Zavlin},
  {Burwitz}, \& {Dodson}}]{pavlov_vela_image}
{Pavlov}, G.~G., {Sanwal}, D., {Garmire}, G.~P., {Zavlin}, V.~E., {Burwitz},
  V., \& {Dodson}, R.~G. 2000, American Astronomical Society Meeting, 196, 0

\bibitem[{{Pavlov} {et~al.}(2001{\natexlab{b}}){Pavlov}, {Zavlin}, {Sanwal},
  {Burwitz}, \& {Garmire}}]{pavlov}
{Pavlov}, G.~G., {Zavlin}, V.~E., {Sanwal}, D., {Burwitz}, V., \& {Garmire},
  G.~P. 2001{\natexlab{b}}, \apjl, 552, L129

\bibitem[{{Radhakrishnan} \& {Deshpande}(2001)}]{desh_rad}
{Radhakrishnan}, V. \& {Deshpande}, A.~A. 2001, A\&A, 379, 551

\bibitem[{Radhakrishnan \& Manchester(1969)}]{rad_glitch}
Radhakrishnan, V. \& Manchester, R. 1969, \nat, 222, 228

\bibitem[{{Shepherd}(1997)}]{difmap}
{Shepherd}, M.~C. 1997, in ASP Conf. Ser. 125: Astronomical Data Analysis
  Software and Systems VI, Vol.~6, 77

\bibitem[{{Spruit} \& {Phinney}(1998)}]{spruit_phinney}
{Spruit}, H.~C. \& {Phinney}, E.~S. 1998, \nat, 393, 139

\bibitem[{Walker \& Chatterjee(1999)}]{vla_memo_23}
Walker, C. \& Chatterjee, S. 1999, Ionospheric corrections using GPS based
  models, VLA Scientific memo~23, Cornell University

\end{thebibliography}

\pagebreak

\begin{figure}
\begin{center}
\epsfig{file=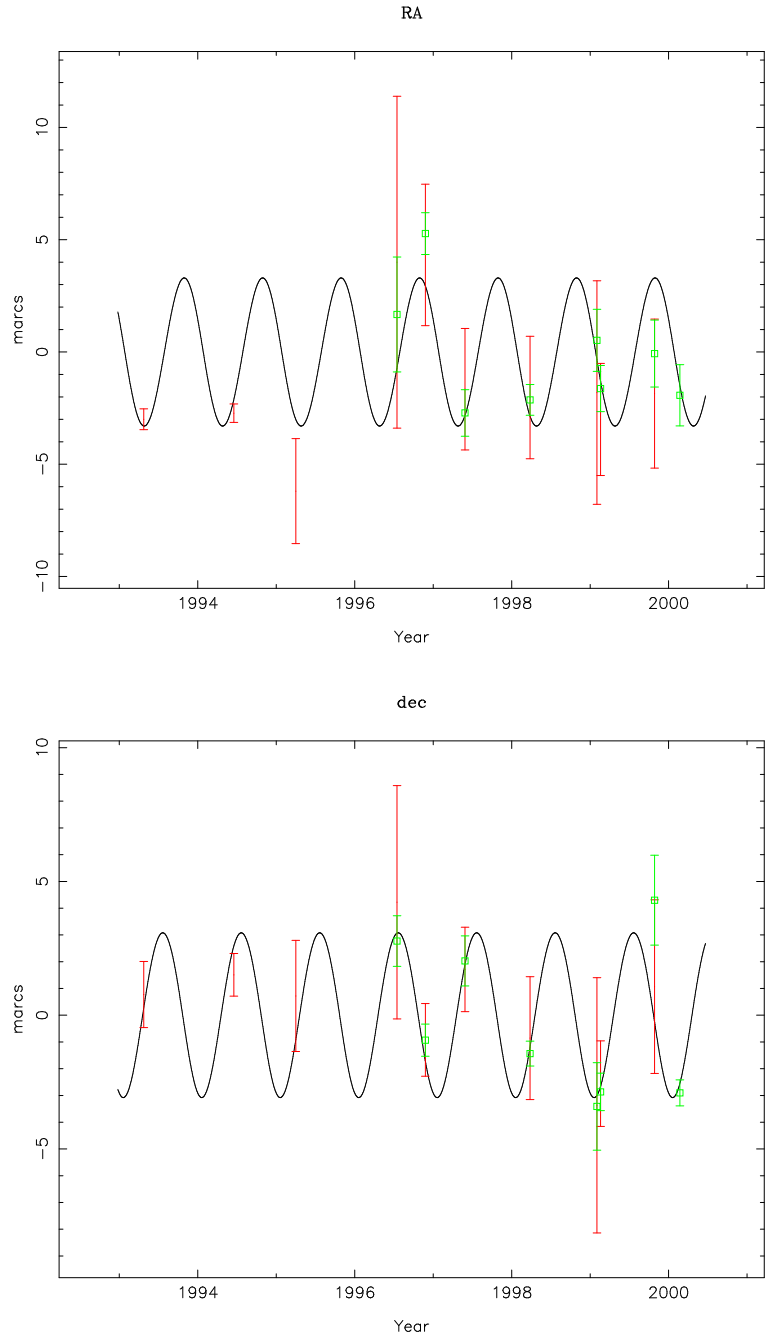,width=10cm,angle=0}
\caption{The residual offsets in the position for the Vela pulsar,
after subtraction of the proper motion, in RA and
Declination. Solutions for 13cm (triangles) and 3cm (boxes)
observations are in shown with error bars in red (13cm) and green
(3cm). The errors are the range of acceptable fits of the to model to
the data, as described in the text, and are significantly smaller for
the 3cm observations, as would be expected.}

\label{fig:offset}
\end{center}
\end{figure}

\pagebreak
\begin{table}
\begin{center}
\begin{scriptsize}
\begin{tabular}{|l|l|l|l|l|}
\hline
Central frequency&Date&Integration&residual\\
MHz              &    &Hours      &mJy\\
\hline
2290& 1993 Apr 24$^\dagger$&13&-\\
2290& 1994 Jun 17$^\dagger$&13&-\\
2290& 1995 Apr 22$^\dagger$&12&-\\
2290&1996 Jul 14&3 & 0.6   \\
8425&1996 Jul 14&3 & 0.9   \\
2290&1996 Nov 23&8 & 0.7   \\
8425&1996 Nov 23&8 & 2.6   \\
2290&1997 May 26&10& 0.7   \\
8425&1997 May 26&7 & 2.5   \\
2290&1998 Mar 26&11& 2.3   \\
8425&1998 Mar 26&11& 1.2   \\
2290&1999 Jan 31&5 & 0.9   \\
8425&1999 Jan 31&5 & 0.9   \\
8425&1999 Feb 17&10& 1.1   \\
2290&1999 Feb 17&10& 1.5   \\
2290&1999 Oct 26&5 & 1.0   \\
8425&1999 Oct 26&5 & 2.0   \\
8417&2000 Feb 21&9 & 1.1   \\
\hline
\end{tabular}
\end{scriptsize}
\end{center}
\caption{Observational details for the data presented here. The image
residual after the subtraction of the point source is also quoted.
$\dagger$Mk III observations.}
\label{tab:observations}
\end{table}

\begin{table}
\begin{center}
\begin{tabular}{ll}
\hline
\multicolumn{2}{c}{In galactic coordinates}\\
\hline
Observed proper motion&$-53.6\pm0.08,-21.9\pm0.06$\\ 
+Correction for the solar motion&$6.7\pm1.6,4.9\pm0.3$\\ 
+Correction for the galactic rotation&$5.35\pm0.3,0.0$\\
\hline
Converted back to RA and Dec&$-38.5\pm1.2,+23.1\pm1.5$\\
\hline
\end{tabular}
\caption{Corrections to the observed Vela proper motion.}
\label{tab:corr}
\end{center}
\end{table}

\begin{table}
\begin{center}
\caption{Vela proper motion determinations. Historic and inaccurate
observations have not been included. These are the proper motions as
seen against the calibrators, i.e. the optical and radio results are
aligned to different reference frames.}
\label{table:Vela proper motion} 
\vspace{0.4cm}
{\scriptsize
\begin{tabular}{llllc}
\hline
\hline
No.& Method     &  $\mu_{\alpha}$ (mas $\mathrm{yr}^{-1}$) & $\mu_{\delta}$ (mas $\mathrm{yr}^{-1}$) & Reference \\
\hline
1&Phase-referencing VLBI                & $-48 \pm 4$&$34 \pm 2$&\citet{bailes}\\
2&Ground-based optical \& HST           & $-41 \pm 3$&$26 \pm 3$&\citet{markwardt}\\
3&Ground-based optical                  & $-47 \pm 3$&$22 \pm 3$&\citet{nasuti}\\
4&HST                                   & $-45 \pm 1$&$26 \pm 1$&\citet{caraveo_pm}\\
5&This Result                           & $-49.61\pm0.06$  & $29.8\pm0.1$&\\
\hline
\end{tabular}
}
\end{center}
\end{table}

\end{document}